\documentclass[aps,prd,preprint,superscriptaddress,preprintnumbers]{revtex4-1}

\pdfoutput=1
\usepackage{mathrsfs}
\usepackage{amsfonts}
\usepackage{amsmath,amssymb,bm,bbm}
\usepackage[colorlinks=true,linkcolor=blue,citecolor=blue,urlcolor=blue]{hyperref}
\usepackage{epsfig}
\usepackage{graphicx}
\usepackage{slashed}
\usepackage{booktabs}
\usepackage{tabu}
\usepackage[dvipsnames]{xcolor}
\usepackage[normalem]{ulem}
\usepackage{tikz}
\usepackage{soul}
\usetikzlibrary{shapes,arrows}
\usetikzlibrary{positioning}
\usepackage{ulem}

\newcommand*\dd{\mathop{}\!\mathrm{d}}

\begin{document}
\preprint{JLAB-THY-23-3950}

\title{Gluon helicity from global analysis of experimental data \\ and lattice QCD Ioffe time distributions}

\author{J.~Karpie}
\affiliation{Jefferson Lab, Newport News, Virginia 23606, USA}

\author{R.~M.~Whitehill}
\affiliation{Department of Physics, Old Dominion University, Norfolk, Virginia 23529, USA}

\author{W.~Melnitchouk}
\affiliation{Jefferson Lab, Newport News, Virginia 23606, USA}

\author{C. Monahan}
\affiliation{Jefferson Lab, Newport News, Virginia 23606, USA}
\affiliation{\mbox{Department of Physics, William \& Mary, Williamsburg, Virginia 23185, USA}}

\author{K.~Orginos}
\affiliation{Jefferson Lab, Newport News, Virginia 23606, USA}
\affiliation{\mbox{Department of Physics, William \& Mary, Williamsburg, Virginia 23185, USA}}

\author{J.-W.~Qiu}
\affiliation{Jefferson Lab, Newport News, Virginia 23606, USA}
\affiliation{\mbox{Department of Physics, William \& Mary, Williamsburg, Virginia 23185, USA}}

\author{D.~G.~Richards}
\affiliation{Jefferson Lab, Newport News, Virginia 23606, USA}

\author{N.~Sato}
\affiliation{Jefferson Lab, Newport News, Virginia 23606, USA}

\author{S.~Zafeiropoulos}
\affiliation{Aix Marseille Univ, Universit\'{e} de Toulon, CNRS, CPT, Marseille, France\\ \vspace*{0.2cm}
{\bf Jefferson Lab Angular Momentum (JAM) and HadStruc Collaborations
\vspace*{0.2cm} }}
	     
\begin{abstract}
We perform a new global analysis of spin-dependent parton distribution functions with the inclusion of Ioffe time pseudo-distributions computed in lattice QCD (LQCD), which are directly sensitive to the gluon helicity distribution, $\Delta g$. These lattice data have an analogous relationship to parton distributions as do experimental cross sections, and can be readily included in global analyses. We focus in particular on the constraining capability of current LQCD data on the sign of $\Delta g$ at intermediate parton momentum fractions $x$, which was recently brought into question by analysis of data in the absence of parton positivity constraints. We find that present LQCD data cannot discriminate between positive and negative $\Delta g$ solutions, although significant changes in the solutions for both the gluon and quark sectors are observed.   
\end{abstract}

\date{\today}
\maketitle

\section{Introduction}

The decomposition of the spin of the proton in terms of its constituent quark and gluon (or parton) degrees of freedom has been the subject of tremendous interest over the last three decades, ever since the discovery by the European Muon Collaboration (EMC)~\cite{EuropeanMuon:1987isl} that the intrinsic spin carried by quarks was only about $\lesssim 10\%-30\%$ of the proton's spin. 
These findings were confirmed by subsequent measurement at CERN, SLAC, DESY, and more recently at Jefferson Lab and RHIC (for reviews, see {\it e.g.}, Refs.~\cite{Lampe:1998eu, Aidala:2012mv}).
Specifically, in the Jaffe-Manohar~\cite{Jaffe:1989jz} decomposition, the proton's spin contributions can be described in terms of the helicity of individual partons and the collective orbital angular momentum originating from quarks and gluons,
\begin{align}
\frac{1}{2} = \frac{1}{2} \Delta \mathbf{\Sigma}(\mu) + \Delta \mathbf{G}(\mu) + \mathbf{L}_{q+g}(\mu) \;.
\end{align}
Here, $\frac{1}{2}\Delta \Sigma(\mu)$ and $\Delta G(\mu)$ denote the net spin contributions from quarks and gluons, respectively, while $L_{q+g}$ represents the corresponding net orbital angular momentum from quarks and gluons. While component in the sum depends on the scale $\mu$ due to renormalization, the sum is a scale-invariant quantity. Utilizing the helicity basis, one can compute the net spin contribution of partons through moments of the helicity-dependent parton distribution functions (hPDFs) as
\begin{align}
    \Delta \mathbf{G}(\mu) 
    &= \int_0^1 \dd x\, \Delta g(x,\mu), 
    \\
    \Delta \mathbf{\Sigma}(\mu) 
    &= \int_0^1 \dd x\, \Delta \Sigma (x,\mu) \\
    &= \sum_q \int_0^1 \dd x\, \big( \Delta q(x,\mu) + \Delta {\bar q}(x,\mu) \big).
\end{align}
where $x$ is the longitudinal light-cone momentum fraction carried by partons relative to their parent proton, and the sum runs over all quark flavors $q=u,d,s,c,b$.

There are several basic considerations that are relevant to point out.
First, hPDFs are of course not directly measurable quantities.
Instead, observables such as double spin asymmetries (DSAs) measured in polarized deep-inelastic scattering (DIS) provide constraints on hPDFs via QCD factorization, which allows for the approximate expression of the measured asymmetries as convolutions of parton-level coefficient functions and hPDFs.
Second, spin asymmetries are unable to impose constraints on hPDFs down to $x=0$, as this would require prohibitively high energies in particle collisions. 
Additionally, standard QCD factorization theorems are only valid provided there is a measurable hard scale $Q$ in the reaction that is large enough for the applicability of perturbative calculations.
This typically limits the lower bounds that experimental data can impose on hPDFs.
The EMC provided constraints on hPDFs down to $x\approx 0.01$ and found that the reconstructed total quark spin $\Delta \mathbf{\Sigma}$ was positive but far too small to account for the proton spin, although constraints were only in the region of $0.01<x<0.5$ with large extrapolation uncertainties. 
At that time, constraints on $\Delta \mathbf{G}$ were also rather nonexistent because the gluon hPDF only enters the DSA at next-to-leading order in perturbative QCD.
Furthermore, constraints on $\Delta g$ via evolution were limited due to the kinematic coverage of the experiments.

With the advent of the RHIC spin experimental program, knowledge about $\Delta g$ began to emerge thanks to measurement of DSAs in inclusive hadron and jet production in polarized proton-proton collisions. 
Using RHIC data~\cite{STAR:2014wox} within a global analysis framework, the DSSV group found the first clearly nonzero signal and a positive gluon hPDF in the region above $x \approx 0.1$~\cite{deFlorian:2014yva}. 
These observations were confirmed in subsequent inclusive jet production data from the STAR~\cite{STAR:2019yqm, STAR:2021mfd, STAR:2021mqa} and PHENIX~\cite{PHENIX:2010aru} collaborations, leading to greater confidence that both the quark and gluon helicity content of the proton were relatively well understood. 
Complementary efforts were also made by the PHENIX collaboration to empirically determine the sign of gluon polarization without relying on global QCD analysis. 
Specifically, in Refs.~\cite{PHENIX:2014axc, PHENIX:2020trf} PHENIX observed an hierarchy of DSAs in hadron production, with $\pi^+ > \pi^0 > \pi^-$, indicating a positive sign for $\Delta g$ based on perturbative QCD arguments.

Recently, the JAM collaboration~\cite{Zhou:2022wzm} revisited the impact of RHIC spin data within a global analysis, with particular focus on the theoretical assumptions that are commonly made in such studies.
Specifically, it was found that parton-level positivity constraints play an important role in determining the sign of $\Delta g$. 
These constraints amount to demanding positivity on the individual helicity components (hPDF$^{\pm}$), such that $g_{\uparrow/\downarrow}(x) > 0$~\footnote{We use ``hPDF'' to denote $\Delta q=q_{\uparrow}-q_{\downarrow}$ and ``hPDF$^\pm$'' for $q_{\uparrow/\downarrow}$, with $q$ labeling a generic parton flavor.}, where
\begin{align}
g_{\uparrow/\downarrow} = \frac{1}{2} \big( g\pm \Delta g \big),
\end{align}
and $g$ is the unpolarized gluon PDF.
Relaxing these constraints in a global analysis reveals a possible second set of solutions in which $\Delta g$ is negative. 
Furthermore, the vast majority of the positive solutions also violate the naive positivity bounds in the very large-$x$ region.
Zhou {\it et al.} \cite{Zhou:2022wzm} showed that all the jet DSA data can be equally well described by the negative $\Delta g$ solutions and by the positive solutions. 
This emphasizes the lack of constraints on hPDFs at large values of $x$ from experimental data, mostly due to the growing statistical uncertainties in DSA measurements at large $x$.

In addition, Whitehill {\it et al.}~\cite{Whitehill:2022mpq} demonstrated that the negative $\Delta g$ solutions can equally well describe the pion DSA data measured by the PHENIX collaboration~\cite{PHENIX:2014axc, PHENIX:2020trf}.
In view of these observations, the PHENIX collaboration recently presented a new analysis of DSAs in isolated prompt-photon data, from which they concluded that the negative $\Delta g$ solutions can be ruled out with a more than $2.8\sigma$ confidence level. 
However, in the PHENIX analysis the unpolarized cross sections that are part of the denominator of the DSA are only describable for photon transverse momentum $p_{\rm T} \gtrsim 10$~GeV (see Fig.~1 of Ref.~\cite{PHENIX:2022lgn}).
This leaves only three out of seven DSA data points above $p_{\rm T}=10$~GeV that are describable within a perturbative QCD framework.
These remaining data points have sufficiently large uncertainties that the disagreement with negative $\Delta g$ solutions would very likely be significantly below the 2.8$\sigma$ confidence level, so the question remains inresolved.

Given the lack of clarity about the sign of the gluon hPDF in the absence of parton positivity constraints, one may be tempted to ask whether it would be prudent to impose such constraints at present until future data can make them redundant. 
Recently Collins {\it et al.}~\cite{Collins:2021vke} pointed out that PDFs in general do not need to be positive definite, even though physical cross sections, as well as individual cross section components in spin asymmetries, must always be positive.
In the DSA $A_{LL} = (\sigma_{+}-\sigma_{-})/(\sigma_{+}+\sigma_{-})$, where $\sigma_{\pm}$ represents the two longitudinal spin configurations of the interacting beams, QCD factorization requires both $\sigma_+$ and $\sigma_-$ to be positive. 
Negative components in PDFs can, in principle, induce negative $\sigma_{\pm}$ contributions, which could be eliminated by imposing the positivity constraints. 
However, other sources, such as large logarithms in fixed-order perturbative calculations or significant power corrections that go beyond standard leading-power treatments, could also bring about such scenarios.
Furthermore, the negative $\Delta g$ found in Ref.~\cite{Zhou:2022wzm} obviously does not violate the positivity of $\sigma_{\pm}$, since all the DSAs are well described and fall within the physical bounds, $|A_{LL}| < 1$. 
Therefore, at present there is no clear data-driven evidence that rules out the negative solutions for $\Delta g$.

One could argue that the phase space coverage of the existing data is not a sufficient condition to accept the negative $\Delta g$ as a physical solution. 
It is of course possible to compute hypothetical observables outside the current experimental reach and find violations of positive cross sections. 
The challenge with this strategy, however, is that it assumes strict validity of factorization and perturbative stability across the entire physical phase space. 
Even if only a conservative region of phase space, where the theoretical framework is expected to operate relatively well, is considered, the lack of empirical evidence that demonstrates that theory can describe a given hypothetical data with the same universal sets of hPDFs describing existing data prevents us from testing universality and the predictive power of the reconstructed hPDFs.

While determining the sign of the gluon polarization will require new experiments at planned facilities, such as those at Jefferson Lab and the future Electron-Ion Collider, an alternative strategy for the present time is to explore off-the-light-cone matrix elements calculable in lattice QCD (LQCD).
A pioneering approach was introduced by Ji~\cite{Ji:2013dva} (for recent reviews see, {\it e.g.}, Refs.~\cite{Ji:2020ena, Liu:2021lke}) within the framework of large momentum effective theory (LaMET), which allows matrix elements of operators with space-like separation to be related to PDFs.
A complementary approach introduced by Radyushkin~\cite{Radyushkin:2017cyf, Orginos:2017kos} allows for this relationship even when the space-like separation is small, removing the formal requirement of large momentum. 
Practically, however, in both approaches a high precision, purely LQCD reconstruction of PDFs is limited by current computational resources, since access to larger momenta and smaller separations incurs greater costs. 
Synergistic activities are currently underway to make use of LQCD data as potential sources of information complementing hadron structure studies where the reach of experiments is limited.
For instance, growing efforts to combine LQCD and experimental data within a global analysis framework have taken place~\cite{Lin:2017snn, Lin:2017stx, JeffersonLabAngularMomentumJAM:2022aix, Gamberg:2022kdb}, which have illustrated that combining information from LQCD with experimental data can lead to stronger constraints on PDFs than those obtained from either LQCD or experimental data alone.

In the context of hPDFs, the quark helicity contribution can be approximately reconstructed from proton matrix elements of the axial current \cite{FlavourLatticeAveragingGroupFLAG:2021npn}, although determining the gluon helicity and orbital angular momentum contributions is more challenging. 
One approach to extracting these quantities requires the computation of matrix elements of local operators which while approximating in the infinite momentum limit are related to $\Delta G$ within the LaMET formalism~\cite{Ji:2013fga, Hatta:2013gta, Zhao:2015kca}.
All of these approaches pose significant difficulties, and currently lattice data only provide weak constraints on the gluon helicity contributions to the proton'sspin~\cite{Yang:2016plb, Khan:2022vot}.

Recently, the HadStruc collaboration has provided new LQCD calculations of matrix elements that have direct sensitivity to $\Delta g$~\cite{HadStruc:2022yaw}. 
In their analysis, it was argued that the negative $\Delta g$ solutions were significantly disfavored by LQCD data.  
Motivated by these findings, in this paper we explore the full extent to which LQCD data can impose constraints on gluon polarization in the proton in terms of QCD factorization approach, and seek a potential resolution regarding its sign.
In Sec.~\ref{sec:latt_data} we review the LQCD calculations of the Ioffe time pseudo-distributions, and summarize the experimental data used in our analysis in Sec.~\ref{sec:exp_data}. 
In Sec.~\ref{sec.analysis} we present the results of the combined analysis of the LQCD and experimental data, offering detailed comparisons of the results before and after the inclusion of the LQCD data. 
Our concluding remarks are found in Sec.~\ref{sec.conclusions}.

\section{Lattice QCD data}\label{sec:latt_data}

In this section we review the LQCD calculations of pseudo-PDFs, as introduced by Radyushkin~\cite{Radyushkin:2017cyf}. This method involves the computation of Lorentz invariant amplitudes (or linear combinations of them) called Ioffe time pseudo-distributions (pseudo-ITDs). The pseudo-ITDs can be matched to the PDFs in the $\overline{\rm MS}$ scheme when the invariant separation between the field operators $z^2$ is sufficiently small. We consider matrix elements of the form~\cite{HadStruc:2021wmh, HadStruc:2022yaw, Fan:2021bcr, Fan:2022kcb, Good:2023ecp}
\begin{equation}
    \widetilde{M}^{\mu \nu ; \alpha \beta}(p,z) 
    = \langle p | F^{\mu\nu}(0) W(0;z) \widetilde{F}^{\alpha\beta}(z) | p \rangle,
\end{equation}
where $F^{\mu\nu}$ and $\widetilde{F}^{\alpha \beta}$ represent the gluon field strength tensor and its dual, with color indices implicitly contracted, and $W$ is a straight Wilson line in the adjoint representation.
In the limit where $z$ is a light-like separation, this matrix element can be used to provide the operator definition for $\Delta g$ that is accessible experimentally. The Lorentz decomposition for the generic matrix element in Ref.~\cite{Balitsky:2021cwr} is rather involved, with fourteen terms that remain after considering the antisymmetry in indices $\mu\leftrightarrow\nu$ and $\alpha\leftrightarrow\beta$, though two constraints exist between multiple terms.
In the operator definition of $\Delta g$, only three of the terms contribute. With space-like separations, it is useful to consider the combination
\begin{subequations}
\begin{align}
\widetilde{M}_{00}(p,z)
&= p_0p_3 \left[ \widetilde{M}^{ti;it}(p,z) + \widetilde{M}^{ij;ji}(p,z) \right]
\\ 
&= \widetilde{\mathcal{M}}(\nu,z^2) 
 + \frac{m^2z^2}{\nu}\mathcal{M}_{pp}(\nu,z^2),
\end{align}
\end{subequations}
where $\nu=p\cdot z$ is the Ioffe time~\cite{Braun:1994jq}, and $i,j$ are spatial directions transverse to $z$. The primary reason to consider such a combination in the space-like separations is that it contains the very same linear combination of the Lorentz invariants that appear in the light-cone case, represented by $\widetilde{\mathcal{M}}$, alongside a power correction term, $\mathcal{M}_{pp}$, proportional to $m^2 z^2/\nu$, where $m$ is the proton mass. It is the $\widetilde{\mathcal{M}}$ term which survives the small-$z^2$ limit and will be related to the parton distributions. The particular combination defining $\widetilde{M}_{00}$ also happens to be multiplicatively renormalizable~\cite{Li:2018tpe}, where the renormalization constant contains an exponential dependence from the Wilson line and a logarithmic dependence determined by the specific choices of indices.

Following the proposal in Ref.~\cite{Balitsky:2021cwr}, we construct the reduced pseudo-ITD as
\begin{equation}
\widetilde{\mathfrak{M}}(\nu,z^2) 
    = \frac{\widetilde{M}_{00}(p,z) 
     \big/ \big[ p_0 p_3 Z_L(z_3/a) \big]}{M_{00}(p=0,z)/m^2} \; .
\label{eq.PITD}
\end{equation}
Note that this quantity is finite in the continuum limit. The combination $M_{00} = M_{ti;it}+M_{ij;ji}$ represents the matrix element for the unpolarized gluon PDF defined in Ref.~\cite{Balitsky:2019krf} which contains the same Wilson line renormalization constant. The factor $Z_L$ cancels the remaining logarithmic ultraviolet divergences. After cancellation of the renormalization constants, the denominator is given by the average gluon momentum fraction $\langle x\rangle_g$. The purpose of this ratio is to construct a calculable observable, finite in the continuum limit, which reduces to the renormalized $\widetilde{\mathcal{M}}$ amplitude in the small-$z^2$ limit where it can be related to the PDFs, or equivalently their Ioffe time distributions. 
The gluon and quark-singlet Ioffe time helicity distributions, ${\cal I}_{\Delta g}$ and $\mathcal{I}_{\Delta\Sigma}$, respectively
\begin{align}
    {\cal I}_{\Delta g} (\nu,\mu^2) 
    &= \int_0^1 \dd x\,x \sin(x\nu)\, \Delta g(x,\mu^2), 
    \\
    \mathcal{I}_{\Delta\Sigma}(\nu,\mu^2)
    &= \int_0^1 \dd x\,x\sin(x\nu)\, \Delta \Sigma(x,\mu^2).
\label{eq:ITD_def}
\end{align}
The matching between the reduced pseudo-ITD and the Ioffe time helicity distributions, is given by~\cite{Balitsky:2021cwr}
\begin{widetext}
\begin{align}
\widetilde{\mathfrak{M}} \left( \nu, z_3^2 \right) \langle x \rangle_g(\mu^2) 
&= {{\cal I}_{\Delta g}(\nu, \mu^2 )} 
- \frac{\alpha_s N_c }{2\pi}\, \int_0^1 \dd u\, {{\cal I}_{\Delta g} (u\nu, \mu^2 ) 
 }  
\nonumber\\
& \quad \times
  \Bigg\{ \log\Big( z_3^2 \mu^2 \frac{e^{2\gamma_E}}{4} \Big)
   \left( \left [\frac{2u^2}{\bar u} + 4u\bar u  \right]_+ - 
   \left( \frac 12  + \frac 43  \frac{\langle x \rangle_\Sigma(\mu^2)}{ \langle x \rangle_g(\mu^2) } \right) \delta( \bar u ) \right) 
\nonumber\\
&\quad\qquad  
+ 4 \left[\frac{u+\log (1-u)}{\bar u}\right]_+ 
- \left[ \frac{1}{ \bar u} - \bar u \right]_+ 
- \frac 12 \delta(\bar u) +2\bar uu \Bigg\}      
\nonumber\\
& - \frac{ \alpha_s C_F}{2\pi}  
\int_0^1 \dd u \, { \mathcal{I}_{\Delta\Sigma}(u \nu,\mu^2) } 
\left( 
  \log \Big(z_3^2 \mu^2 \frac{ e^{2 \gamma_E} }{ 4 } \Big) \widetilde {\cal B}_{gq} (u) + 2\bar uu 
\right) 
\nonumber\\
& + O(m^2 z^2) + O(\Lambda_{QCD}^2 z^2)\,,
\label{eq.ITD}
\end{align}
\end{widetext}
where $\langle x \rangle_\Sigma(\mu^2)$ is the average momentum fraction of the unpolarized quark singlet distribution, $\bar u = 1-u$, and $\widetilde{\cal B}_{gq} (u)=1 - \bar{u}^2$ is the quark-gluon mixing term of the evolution kernel. Note that the factorization is only valid in the limit where $\mathcal{M}_{pp}$ does not contributing to $\widetilde{\mathfrak{M}}$. As will be discussed later, multiple ways were tested in Ref.~\cite{HadStruc:2022yaw} to remove its contribution.

The presence of the structure-dependent momentum fractions 
$\langle x \rangle_g(\mu^2)$ and 
$\langle x \rangle_\Sigma(\mu^2)$ 
in the matching relation is atypical in the analogous factorization of cross sections. It appears entirely due to the evolution of the momentum fraction on the left hand side of Eq.~(\ref{eq.ITD}), which must be included due to the normalization of $\widetilde{\mathfrak{M}}$.
This normalization is convenient for two reasons. Not only does the exponential renormalization of the Wilson line cancel, but it does so in such a way so as to cancel the statistical fluctuations of $\widetilde{M}_{00}$ and $M_{00}$, which are highly correlated.

Note that Eq.~(\ref{eq.ITD}), as all factorization relationships, is valid up to the power correction terms, which in this case are $O(z^2)$. However, it was found~\cite{HadStruc:2022yaw} that these corrections were actually the dominant contribution to the matrix element. To address this, two approaches were used to remove such contributions: one approach involved modeling the two terms in Eq.~(\ref{eq.PITD}) with polynomials in $\nu$, while the other involved subtracting the rest frame matrix element which is exclusively given by the contaminating power correction term. 
The rest frame subtracted data were found to be consistent with the model of $\widetilde{\mathcal{M}}$ from the first approach, giving confidence that both approaches provide consistent results. This agreement implies that the residual contamination from the power corrections has been significantly reduced relative to the overall uncertainty on the leading power contribution in Eq.~(\ref{eq.PITD}). In this study we will apply the factorization \eqref{eq.ITD} to relate a model PDF to the rest frame subtracted data.
Furthermore, in Ref.~\cite{Khan:2022vot} both terms were modeled with a neural network functional form, showing relatively good agreement with the polynomial approach (see Fig.~3 in Ref.~\cite{Khan:2022vot}).

Calculations in LQCD are limited in the maximum momentum that a hadron can carry. Large momentum calculations are plagued by polynomially growing lattice systematic errors and, worse, exponentially growing statistical noise. This issue limits calculations to momenta $|p|\lesssim3$~GeV. With a limited range of $p$, or equivalently $\nu$, the pseudo-ITD cannot constrain the full $x$ region of the PDF. It has been shown~\cite{Karpie:2019eiq} that increasing the range of $\nu$ allows for more accurate reproduction in the low-$x$ region. Even with only $\nu<10$, the hPDF can be determined accurately for $x \gtrsim 0.25$. In Ref.~\cite{JeffersonLabAngularMomentumJAM:2022aix} this feature was exploited by combining experimental results, sensitive to low $x$, and lattice results, sensitive to large $x$, to obtain stronger constraints on the unpolarized quark PDF in the pion. It is the goal of this study to explore whether the polarized gluon pseudo-ITD has sufficient constraining power to discriminate between the sign of $\Delta g$ in the large-$x$ region.

In our study, we include LQCD data that were generated on 1901 configurations of an ensemble with (2+1)-dynamical clover Wilson fermions with stout-link smearing and tree-level tadpole-improved gauge action with a lattice volume $32^3 \times 64$. The lattice spacing is $a=0.094(1)$~fm, determined using the $w_0$ scale~\cite{BMW:2012hcm}, and the pion mass is $m_\pi=358(3)$~MeV, respectively. While the quarks, and thereby pions, have unphysically large masses, this is not expected to be a dominant systematic error for gluon matrix elements compared to discretization effects and other systematic uncertainties. The two-point correlation functions are constructed using the distillation approach~\cite{HadronSpectrum:2009krc} with sources on all possible time slices. Wilson gradient flow~\cite{Luscher:2010iy} was used to control statistical errors, with an extrapolation to zero flow time. The scale dependence entering in Eq.~(\ref{eq.ITD}) is set as 
\begin{align}
    \mu^2={\rm max}\left[\, m_c^2,\ \frac{4}{e^{2\gamma_E} z_3^2}\, \right],
    \label{eq.scale}
\end{align}
where $z_3=a n$ is the space-like separation of the gluon fields, expressed in terms of the lattice spacing $a$, and $n$ is an integer. We choose this scale to optimize the perturbative expression in Eq.~(\ref{eq.ITD}) to remove the logarithmic contributions.

\section{Experimental data}\label{sec:exp_data}

From the experimental side, in the current analysis we restrict ourselves to using only spin observables that are directly sensitive to hPDFs, in contrast to the recent JAM analysis~\cite{Cocuzza:2022jye}, where PDFs, hPDFs and fragmentation functions were all simultaneously extracted from data. Here, we summarize all the experimental data in our analysis:
\begin{itemize}

    \item \underline{DSAs in inclusive DIS}: We include all data from fixed-target experiments conducted by the EMC~\cite{EuropeanMuon:1989yki}, SMC~\cite{SpinMuon:1998eqa, SpinMuon:1999udj}, COMPASS~\cite{COMPASS:2010wkz, COMPASS:2006mhr, COMPASS:2015mhb}, SLAC~\cite{Baum:1983ha, E143:1998hbs, E155:2000qdr, E155:1999pwm, E142:1996thl, E154:1997xfa}, and HERMES~\cite{HERMES:1997hjr, HERMES:2006jyl} collaborations. We apply identical cuts on $W^2$ and $Q^2$ as those used for unpolarized DIS data~\cite{Sato:2016tuz, Ethier:2017zbq}. Whenever available, we use DSAs rather than the reconstructed $g_1$ structure function to ensure consistent propagation of uncertainties include those from PDFs entering in the denominator of the asymmetries. To ensure that the asymmetries are dominated by the leading twist $g_1$ structure function, with negligible contributions from $g_2$, we impose constraints on the four-momentum transfer squared $Q^2 > m_c^2$, and the hadronic final state masses $W^2 > 10$~GeV$^2$.

    \item \underline{DSAs in semi-inclusive DIS (SIDIS)}: With the same cuts as in the inclusive DIS case, we include pion, kaon, and unidentified hadron SIDIS measurements on polarized proton, deuteron, and $^3$He targets from  HERMES~\cite{HERMES:2004zsh,HERMES:1999uyx}, COMPASS~\cite{COMPASS:2009kiy, COMPASS:2010hwr} and SMC~\cite{SpinMuon:1997yns}. The fragmentation variable $z$ is restricted to the range $0.2 < z < 0.8$ to ensure the applicability of the leading-power formalism and avoid hadron mass corrections and threshold effects~\cite{Moffat:2021dji, Guerrero:2017yvf}.

    \item \underline{DSAs in inclusive jet production in polarized $pp$ collisions}: We include DSAs from the STAR~\cite{STAR:2006opb, STAR:2012hth, STAR:2014wox,  STAR:2019yqm, STAR:2021mqa} and PHENIX~\cite{PHENIX:2010aru} collaborations at RHIC. The $p_{\rm T}$ range is restricted to be the same as the minimum $p_{\rm T}$ for which the corresponding unpolarized jet data are describable~\cite{Zhou:2022wzm}. This ensures a faithful description of the denominator in the asymmetries.

\end{itemize}

For all the observables we employ a next-to-leading order framework for the parton level cross sections and asymmetries. The scale settings for DIS and SIDIS are all set equal to the scale of the virtual photon. In the case of jet data, we use the scale settings equal to $\frac12 p_{\rm T}$, which generally yields the best agreement for both unpolarized and polarized data.

\section{Global analysis with LQCD data}
\label{sec.analysis}

Our numerical approach to infer hPDFs in the combined analysis follows the same Monte Carlo strategy as in previous JAM analyses~\cite{Zhou:2022wzm, Cocuzza:2022jye}. Specifically, we employ a data resampling technique where pseudodata are generated by sampling the original data with Gaussian distributions within the uncertainties. In the case of LQCD pseudo-ITD data, we utilize the full covariance matrix for generating pseudodata. For each set of pseudodata, we optimize the hPDF parameters while assigning prior parameters for the PDFs and fragmentation functions from an earlier JAM analysis~\cite{Cocuzza:2022jye}. The resulting ensemble of optimized hPDFs represents the posterior density of the combined LQCD+experimental global analysis.

\begin{table}[t]
    \caption{Labels of the data sets in Figs.~\ref{f.rchi2} and~\ref{f.zscore}. Data sets 1--18 are from polarized DIS, data sets 19--45 are from polarized SIDIS, data sets 46--53 are from jets in polarized $pp$ collisions, and data set 54 is from LQCD. Further information on the lattice and experimental data can be found in Secs.~\ref{sec:latt_data} and \ref{sec:exp_data}, respectively.\\}
    \label{tab:data_names}
    \centering
\scriptsize
    \begin{tabular}{r l |  r l} \hline
(1)& HERMES~$A_1$($n$)~\cite{HERMES:1997hjr, HERMES:2006jyl}~&
~(2)& EMC~$A_1$($p$)~\cite{EuropeanMuon:1989yki}\\
(3)& SMC~$A_1$($p$)~\cite{SpinMuon:1998eqa, SpinMuon:1999udj}~&
~(4)& SLAC(E154)~$A_{\parallel}$($h$)~\cite{Baum:1983ha, E143:1998hbs, E155:2000qdr, E155:1999pwm, E142:1996thl, E154:1997xfa}\\
(5)& HERMES~$A_{\parallel}$($p$)~\cite{HERMES:1997hjr, HERMES:2006jyl}~&
~(6)& SMC~$A_1$($d$)~\cite{SpinMuon:1998eqa, SpinMuon:1999udj}\\
(7)& COMPASS~$A_1$($d$)~\cite{COMPASS:2010wkz, COMPASS:2006mhr, COMPASS:2015mhb}~&
~(8)& SLACE80E130~$A_{\parallel}$($p$)~\cite{Baum:1983ha, E143:1998hbs, E155:2000qdr, E155:1999pwm, E142:1996thl, E154:1997xfa}\\
(9)& SLAC(E143)~$A_{\parallel}$($p$)~\cite{Baum:1983ha, E143:1998hbs, E155:2000qdr, E155:1999pwm, E142:1996thl, E154:1997xfa}~&
~(10)& COMPASS~$A_1$($p$)~\cite{COMPASS:2010wkz, COMPASS:2006mhr, COMPASS:2015mhb}\\
(11)& SLAC(E155)~$A_{\parallel}$($d$)~\cite{Baum:1983ha, E143:1998hbs, E155:2000qdr, E155:1999pwm, E142:1996thl, E154:1997xfa}~&
~(12)& SLAC(E142)~$A_1$($h$)~\cite{Baum:1983ha, E143:1998hbs, E155:2000qdr, E155:1999pwm, E142:1996thl, E154:1997xfa}\\
(13)& HERMES~$A_{\parallel}$($d$)~\cite{HERMES:1997hjr, HERMES:2006jyl}~&
~(14)& SLAC(E143)~$A_{\parallel}$($d$)~\cite{Baum:1983ha, E143:1998hbs, E155:2000qdr, E155:1999pwm, E142:1996thl, E154:1997xfa}\\
(15)& COMPASS~$A_1$($p$)~\cite{COMPASS:2010wkz, COMPASS:2006mhr, COMPASS:2015mhb}~&
~(16)& SMC~$A_1$($p$)~\cite{SpinMuon:1998eqa, SpinMuon:1999udj}\\
(17)& SLAC(E155)~$A_{\parallel}$($p$)~\cite{Baum:1983ha, E143:1998hbs, E155:2000qdr, E155:1999pwm, E142:1996thl, E154:1997xfa}~&
~(18)& SMC~$A_1$($d$)~\cite{SpinMuon:1998eqa, SpinMuon:1999udj}\\
(19)& COMPASS~$A_1$($p$, $K^+$)~\cite{COMPASS:2009kiy, COMPASS:2010hwr}~&
~(20)& COMPASS~$A_1$($p$, $K^-$)~\cite{COMPASS:2009kiy, COMPASS:2010hwr}\\
(21)& HERMES~$A_1$($p$, $\pi^-$)~\cite{HERMES:2004zsh,HERMES:1999uyx}~&
~(22)& COMPASS~$A_1$($d$, $\pi^+$)~\cite{COMPASS:2009kiy, COMPASS:2010hwr}\\
(23)& COMPASS~$A_1$($d$, $K^+$)~\cite{COMPASS:2009kiy, COMPASS:2010hwr}~&
~(24)& COMPASS~$A_1$($d$, $\pi^-$)~\cite{COMPASS:2009kiy, COMPASS:2010hwr}\\
(25)& HERMES~$A_1$($h$, $h^-$)~\cite{HERMES:2004zsh,HERMES:1999uyx}~&
~(26)& HERMES~$A_1$($p$, $h^-$)~\cite{HERMES:2004zsh,HERMES:1999uyx}\\
(27)& HERMES~$A_1$($d$, $K^++K^-$)~\cite{HERMES:2004zsh,HERMES:1999uyx}~&
~(28)& SMC~$A_1$($d$, $h^+$)~\cite{SpinMuon:1997yns}\\
(29)& HERMES~$A_1$($h$, $h^+$)~\cite{HERMES:2004zsh,HERMES:1999uyx}~&
~(30)& HERMES~$A_1$($p$, $h^+$)~\cite{HERMES:2004zsh,HERMES:1999uyx}\\
(31)& COMPASS~$A_1$($d$, $K^-$)~\cite{COMPASS:2009kiy, COMPASS:2010hwr}~&
~(32)& COMPASS~$A_1$($d$, $h^+$)~\cite{COMPASS:2009kiy, COMPASS:2010hwr}\\
(33)& HERMES~$A_1$($d$, $h^-$)~\cite{HERMES:2004zsh,HERMES:1999uyx}~&
~(34)& HERMES~$A_1$($d$, $K^-$)~\cite{HERMES:2004zsh,HERMES:1999uyx}\\
(35)& HERMES~$A_1$($d$, $K^+$)~\cite{HERMES:2004zsh,HERMES:1999uyx}~&
~(36)& COMPASS~$A_1$($d$, $h^-$)~\cite{COMPASS:2009kiy, COMPASS:2010hwr}\\
(37)& COMPASS~$A_1$($p$, $\pi^+$)~\cite{COMPASS:2009kiy, COMPASS:2010hwr}~&
~(38)& SMC~$A_1$($p$, $h^-$)~\cite{SpinMuon:1997yns}\\
(39)& SMC~$A_1$($d$, $h^-$)~\cite{SpinMuon:1997yns}~&
~(40)& COMPASS~$A_1$($p$, $\pi^-$)~\cite{COMPASS:2009kiy, COMPASS:2010hwr}\\
(41)& HERMES~$A_1$($d$, $\pi^+$)~\cite{HERMES:2004zsh,HERMES:1999uyx}~&
~(42)& HERMES~$A_1$($p$, $\pi^+$)~\cite{HERMES:2004zsh,HERMES:1999uyx}\\
(43)& HERMES~$A_1$($d$, $h^+$)~\cite{HERMES:2004zsh,HERMES:1999uyx}~&
~(44)& SMC~$A_1$($p$, $h^+$)~\cite{SpinMuon:1997yns}\\
(45)& HERMES~$A_1$($d$, $\pi^-$)~\cite{HERMES:2004zsh,HERMES:1999uyx}~&
~(46)& PHENIX~$A_{LL}$~$\sqrt{s}=200$~{\rm GeV} (2005)~\cite{PHENIX:2010aru}\\
(47)& STAR~$A_{LL}$~$\sqrt{s}=200$~{\rm GeV} (2015)~\cite{STAR:2006opb, STAR:2012hth, STAR:2014wox,  STAR:2019yqm, STAR:2021mqa}~&
~(48)& STAR~$A_{LL}$~$\sqrt{s}=200$~{\rm GeV} (2005)~\cite{STAR:2006opb, STAR:2012hth, STAR:2014wox,  STAR:2019yqm, STAR:2021mqa}\\
(49)& STAR~$A_{LL}$~$\sqrt{s}=510$~{\rm GeV} (2012)~\cite{STAR:2006opb, STAR:2012hth, STAR:2014wox,  STAR:2019yqm, STAR:2021mqa}~&
~(50)& STAR~$A_{LL}$~$\sqrt{s}=200$~{\rm GeV} (2003)~\cite{STAR:2006opb, STAR:2012hth, STAR:2014wox,  STAR:2019yqm, STAR:2021mqa}\\
(51)& STAR~$A_{LL}$~$\sqrt{s}=200$~{\rm GeV} (2006)~\cite{STAR:2006opb, STAR:2012hth, STAR:2014wox,  STAR:2019yqm, STAR:2021mqa}~&
~(52)& STAR~$A_{LL}$~$\sqrt{s}=510$~{\rm GeV} (2013)~\cite{STAR:2006opb, STAR:2012hth, STAR:2014wox,  STAR:2019yqm, STAR:2021mqa}\\
(53)& STAR~$A_{LL}$~$\sqrt{s}=200$~{\rm GeV} (2009)~\cite{STAR:2006opb, STAR:2012hth, STAR:2014wox,  STAR:2019yqm, STAR:2021mqa}~&
~(54)& HadStruc~$\widetilde{\mathfrak{M}}(p)$~\cite{HadStruc:2022yaw} \\
\hline
    \end{tabular}
\normalsize
\end{table}

\begin{figure*}
    \centering
    \includegraphics[width=\textwidth]{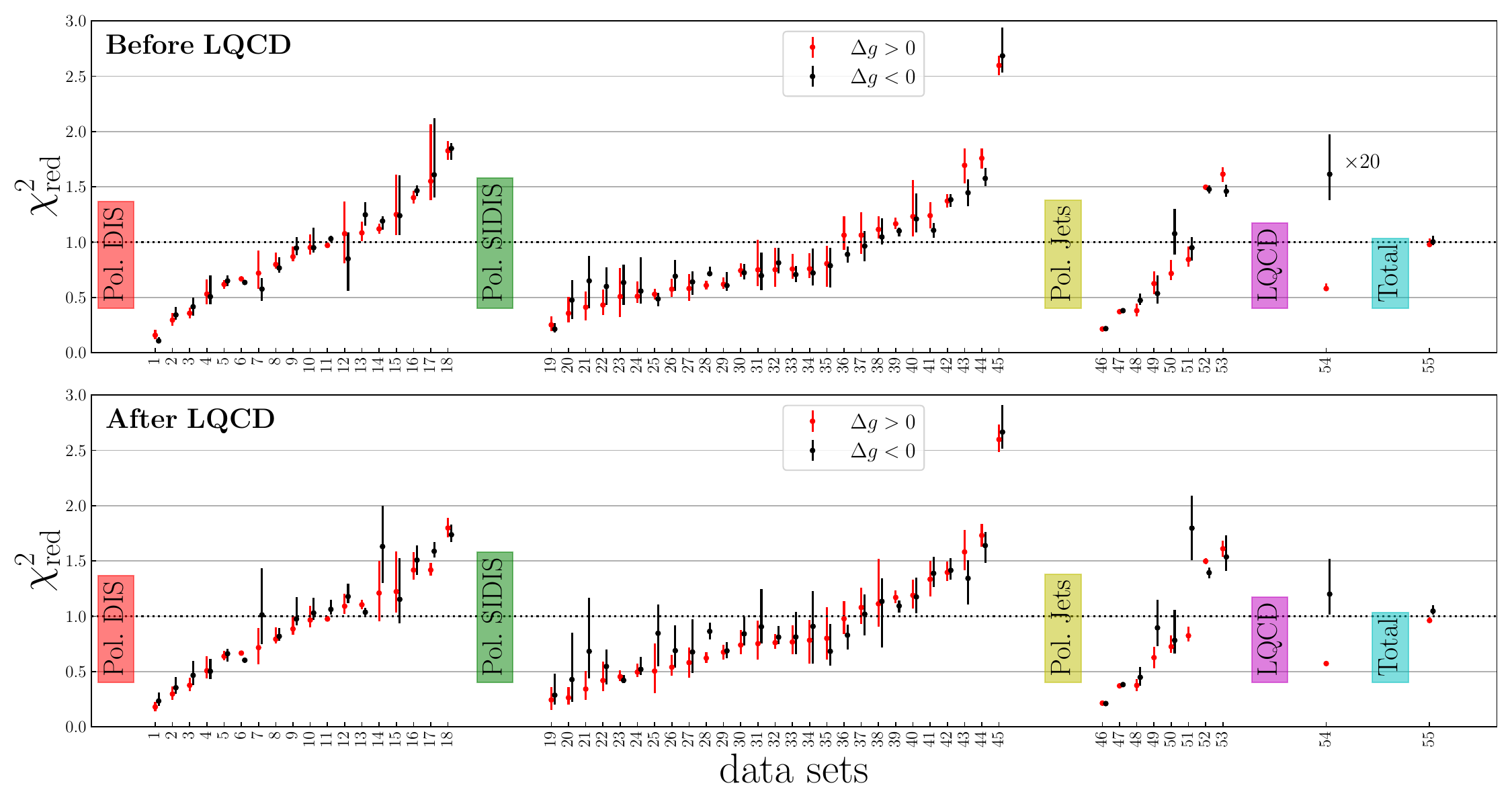}
    \caption{
    Distribution of reduced $\chi^2$ values from PDF replicas per data set ordered by types of data and increasing values of $\chi^2_{\rm red}$ from positive gluon polarization. The negative gluon case for entry 54 has been rescaled by a factor of 20 to fit within the graph in the upper panel, and it corresponds to predictions for LQCD data using the hPDFs that were inferred without the LQCD data. The data set labels are given in Table~\ref{tab:data_names}.
    }
    \label{f.rchi2}
\end{figure*}

After collecting all the hPDF Monte Carlo samples, including the LQCD data, we find that the negative $\Delta g$ solutions still persist, although with significant changes in their shape. To assess the significance of the results, we first discuss the quality of the agreement between the data and theory. Figure~\ref{f.rchi2} displays the reduced $\chi^2$ for the individual data sets, defined as $\chi^2_{\rm red} = \chi^2/N$, where $N$ represents the number of points. We present results both before (from Ref.~\cite{Cocuzza:2022jye}) and after the inclusion of LQCD data. The results are separated by different types of data sets and arranged in increasing order of $\chi^2_{\rm red}$. We tabulate the data sets and their labels in Table~\ref{tab:data_names}. In addition, we categorize the results based on the sign of $\Delta g$ to illustrate the global agreement of the negative solutions in the absence of positivity constraints.

\begin{figure*}
    \centering
    \includegraphics[width=\textwidth]{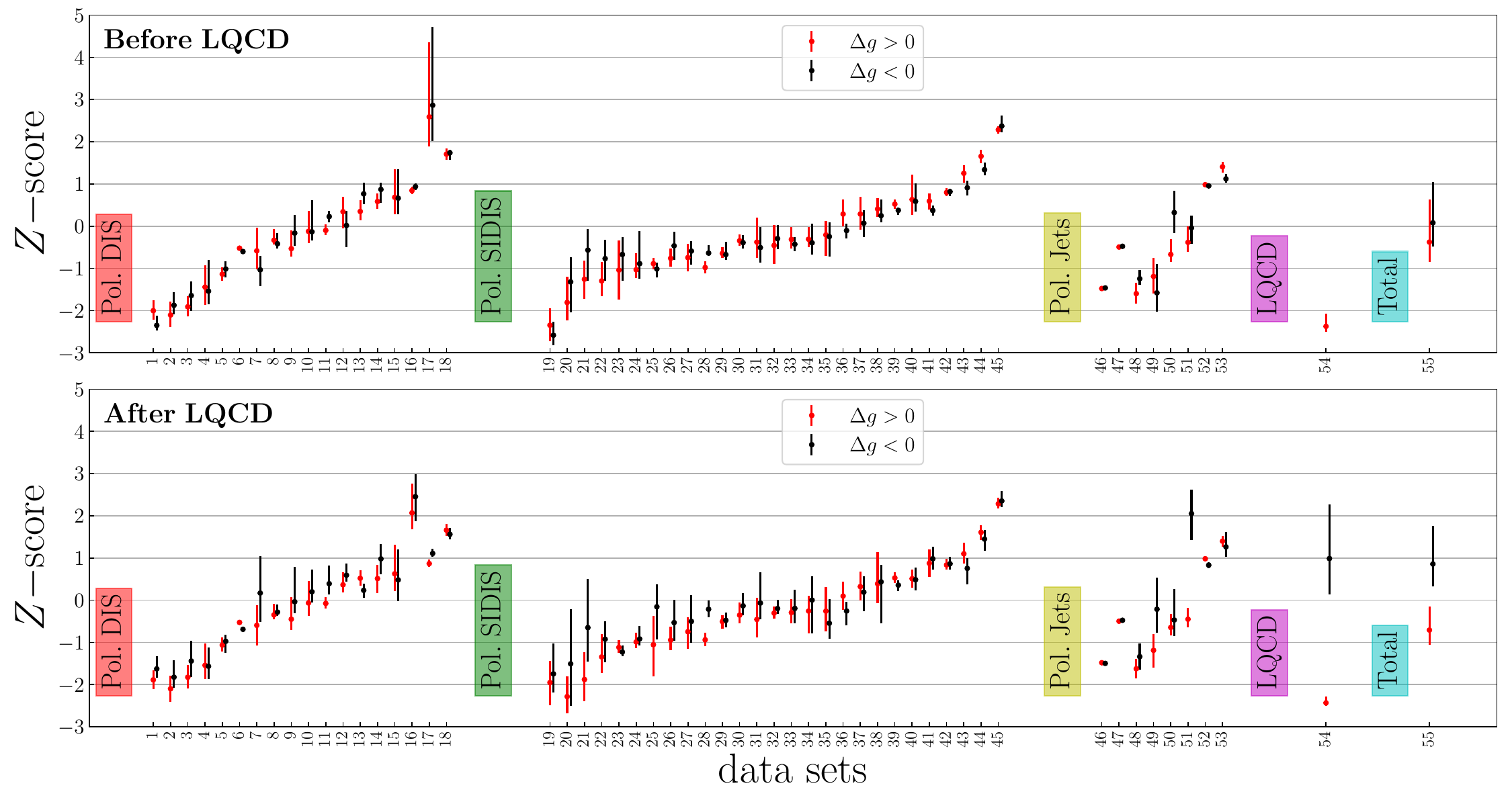}
    \caption{Distribution of $Z-$scores values from PDF replicas per data set ordered by types of data and increasing values of $\chi^2_{\rm red}$ from positive gluon polarization. In the case of the negative gluon for entry 54, the upper panel indicates an approximately infinite value, suggesting an extremely poor prediction of the LQCD data using the prior hPDFs that were inferred from experimental data alone.
    The data set labels are given in Table~\ref{tab:data_names}.}
    \label{f.zscore}
\end{figure*}

In Fig.~\ref{f.zscore}, we provide standardized $Z$-scores based on the Gaussian hypothesis, computed as $Z=\sqrt{2}\, \mathrm{erf}^{-1}(1-2p)$, where $p$ is the $p$-value estimated from a $\chi^2$ distribution with $N$ as the degrees of freedom. This allows us to assess the statistical significance of the reduced $\chi^2$ values and diagnose instances where the $\chi^2$ values deviate from the ideal value of unity. In both figures, the error bars indicate the 50\% percentiles and their neighborhoods of $\pm 1\sigma$ percentiles.

\begin{figure*}[h!]
    \centering
    \includegraphics[width=\textwidth]{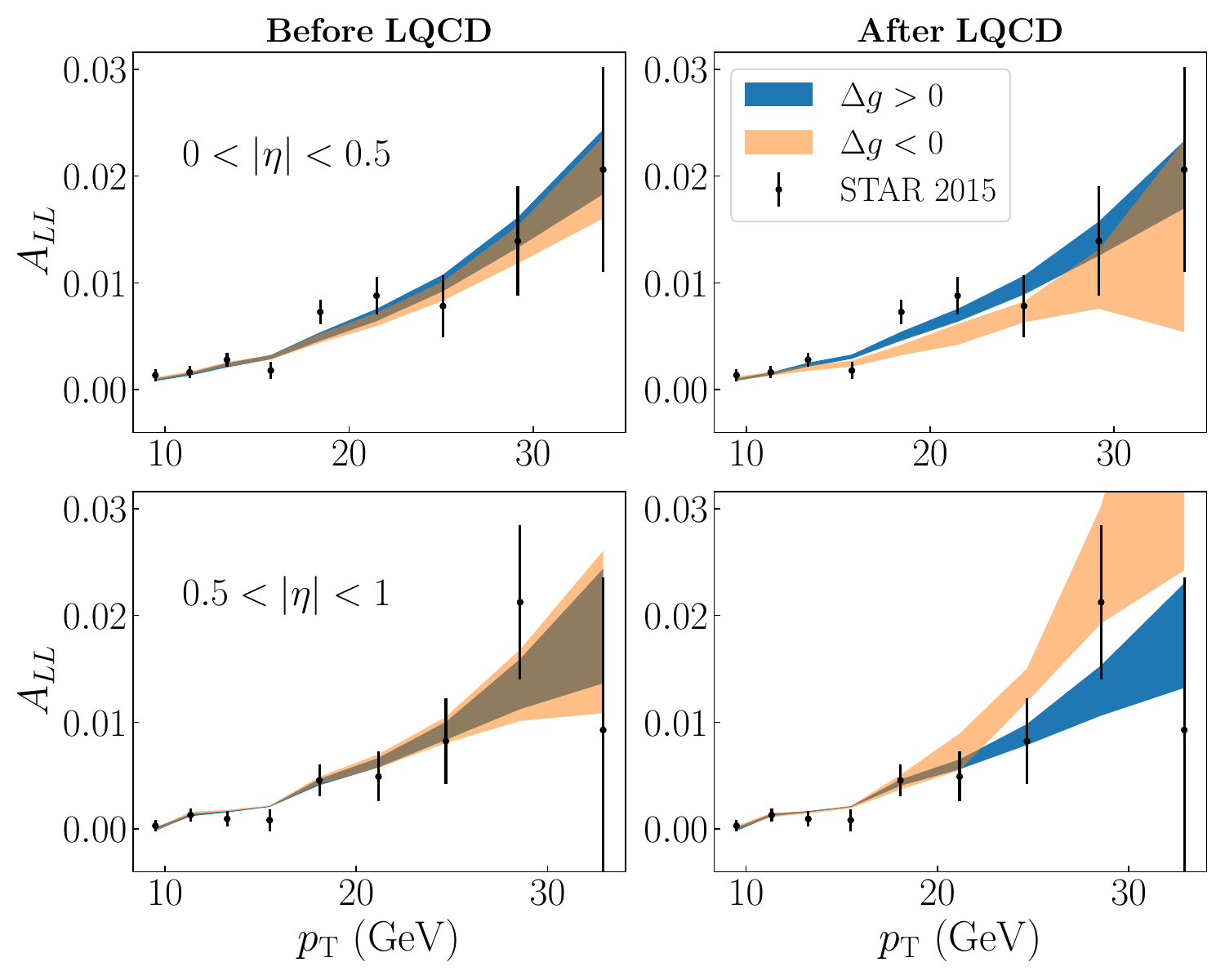}
    \caption{DSAs from STAR collaboration. The figure compares theory and data before and after the inclusion of LQCD data.}
    \label{f.STAR}
\end{figure*}

Prior to the inclusion of LQCD data, most of the experimental data sets exhibit relatively good agreement with the theory, with $Z$-scores confined within 1$\sigma$ in most cases, regardless of the sign of $\Delta g$. However, the LQCD data shows a significant tension for the negative $\Delta g$ solutions. After the inclusion of the LQCD data, one finds the same agreement across most of the data sets as before, with a possible exception in one of the polarized jet data sets labeled as data set ``51" in Fig.~\ref{f.zscore}. This data set corresponds to DSAs in polarized jets from the STAR collaboration. To examine this, in Fig.~\ref{f.STAR}, we show the data and theory comparisons. The inclusion of the LQCD data forces the negative solutions to deviate further from a few $A_{LL}$ data points around $p_{\rm T}\sim 20$ GeV at the $0<|\eta|<0.5$ bin, causing an increase in the $Z$-score from $< 1\sigma$ to $2\sigma$, which is however not statistically significant.

Note that in principle it is possible to obtain physical $|A_{LL}| < 1$ DSAs with $\sigma_+$ and $\sigma_-$ both negative. However, this would imply that the spin-averaged cross sections, proportional to $\sigma_+ + \sigma_-$, would also be negative. Since we agree with the unpolarized cross section data, including at RHIC kinematics, this scenario can be ruled out in our analysis.

\begin{figure*}[h!]
    \centering
    \includegraphics[width=\textwidth]{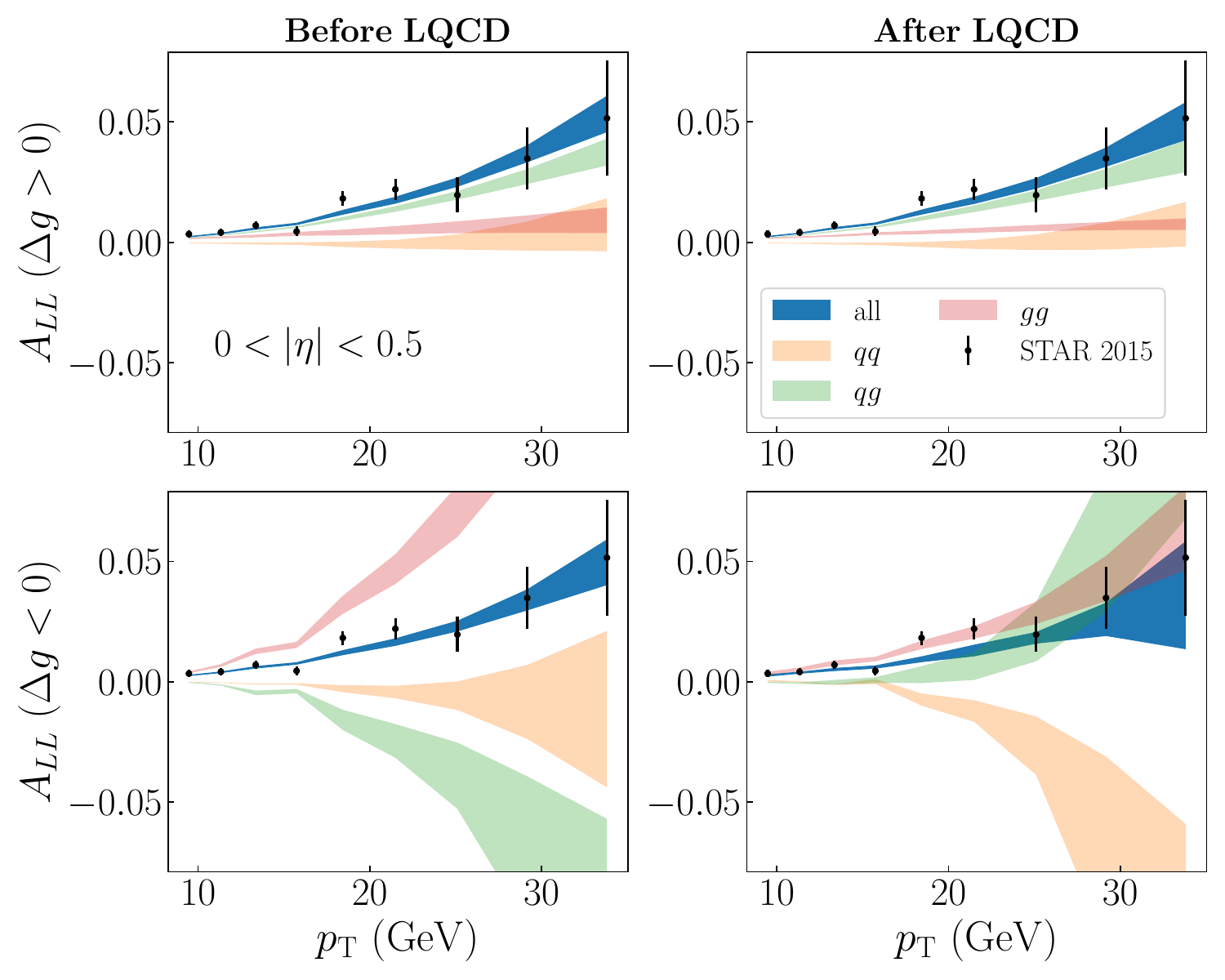}
    \caption{Quark and gluon sub-processes contributing to DSAs data from the STAR collaboration. The figure compares theory and data before and after the inclusion of LQCD data.}
    \label{f.qg-channels}
\end{figure*}

Taking the same polarized jet data set 51, it is instructive to decompose its numerator into the three possible partonic subprocesses: $qq$, $qg$, and $gg$, to understand the role of the linear term with $\Delta g$ that can discriminate its sign. This is shown in Fig.~\ref{f.qg-channels} for the $0<|\eta|<0.5$ bin for the two solutions of $\Delta g$ and compares the results before and after the inclusion of LQCD data. 
In the case of $\Delta g>0$, it is clear that the linear contribution $qg$ is the leading subprocess of the DSAs at larger values of $p_{\rm T}$ relative to the other subprocesses, and the inclusion of LQCD data does not significantly alter the relative contributions of the subprocesses. In contrast, prior to the inclusion of LQCD data, the negative $\Delta g$ solutions enhance the role of the $gg$ channel at the expense of making the $qg$ channel more negative in order to balance out the relative contributions to the DSAs and describe the data. This situation changes with the the inclusion of LQCD data where the $qg$ and $gg$ channels contribute positively at larger values of $p_{\rm T}$ at the expense of turning the $qq$ channel negative. This means that the quark hPDFs have undergone changes at large $x$, despite the fact that all the DSAs from DIS up to $x \sim 0.66$ considered in this analysis are well described.

\begin{figure*}
    \centering
    \includegraphics[width=\textwidth]{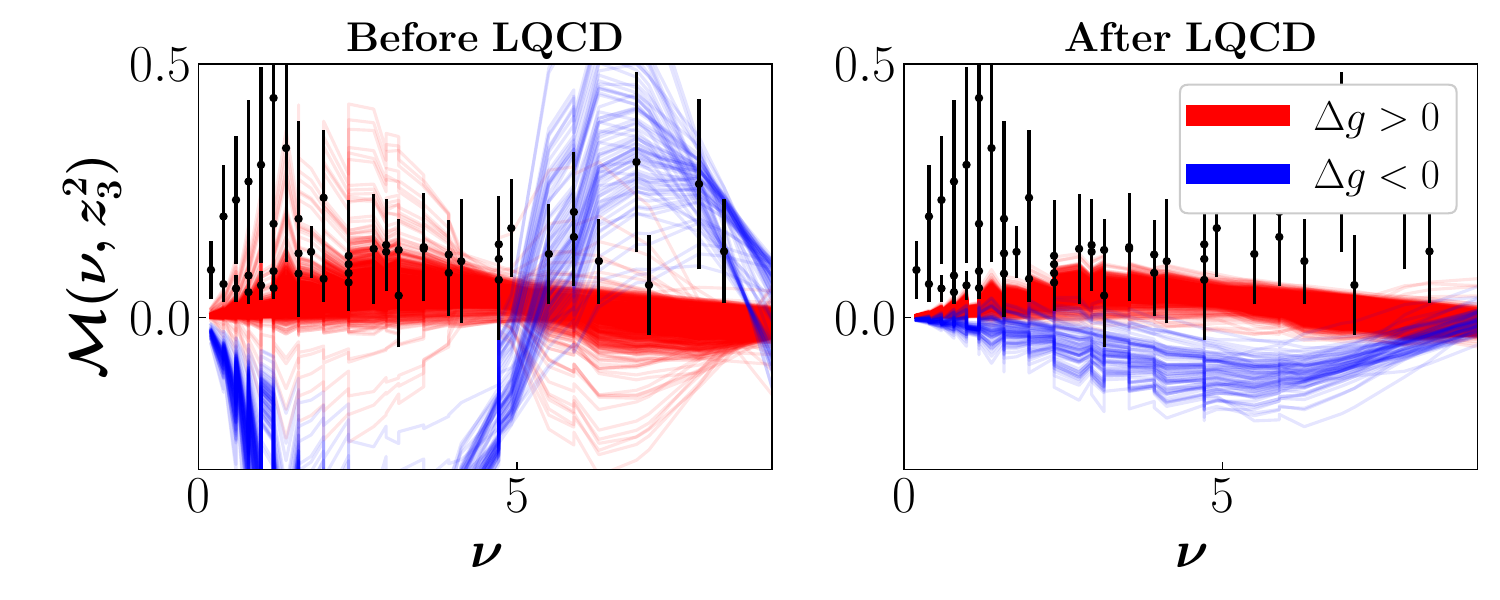}
\caption{Comparison of LQCD data with Eq.~\ref{eq.ITD} using hPDF from global analysis without LQCD data (left) and with LQCD data (right).  }
    \label{f.lqcd-pheno}
\end{figure*}

We also find that the inclusion of the LQCD data admits negative solutions for $\Delta g$ that can describe the LQCD data relatively well, with $Z$-scores ranging from 1-3$\sigma$, which, in turn, prevents the complete elimination of the negative solutions from the posterior distribution. To understand the situation, in Fig.~\ref{f.lqcd-pheno}, we display the lattice data as a function of the Ioffe time $\nu$. The data points are available at different values of $z_3^2$ for each value of $\nu$, which requires us to use different values for the scale settings in Eq.~\ref{eq.ITD}. 
The calculations of Eq.~\ref{eq.ITD} are performed at discrete values of $\nu$ and $z_3^2$, and we have linearly connected the points to show the trends for the positive (red) and negative (blue) $\Delta g$ solutions. Prior to the inclusion of the LQCD data, the positive solutions exhibit relatively good agreement with the data, while the negative solutions display a peculiar oscillatory behavior that is inconsistent with the data. This inconsistency is particularly noticeable in the lower $\nu$ regions, where LQCD calculations are expected to be more reliable.  After the inclusion of the LQCD data, the variance of the positive solutions decreases, indicating a level of constraint on the hPDFs. However, the negative solutions persist, albeit with a shape that exhibits fewer oscillations. These two solution sets clearly have distinctive signs for $\widetilde{\mathfrak{M}}$. Since the majority of the LQCD data is positive, the negative solutions are disfavored. From a global analysis perspective, these negative solutions do not disappear entirely due to the contribution of the $\chi^2$ function from the LQCD data, which includes a covariance matrix with non-zero off-diagonal components not included in Fig.~\ref{f.lqcd-pheno}. When considering the full covariance matrix of the LQCD data, one finds that the negative solutions agree within approximately 1$\sigma$ confidence level, as shown in Fig.~\ref{f.zscore}.

\begin{figure*}[h!]
    \centering
    \includegraphics[width=\textwidth]{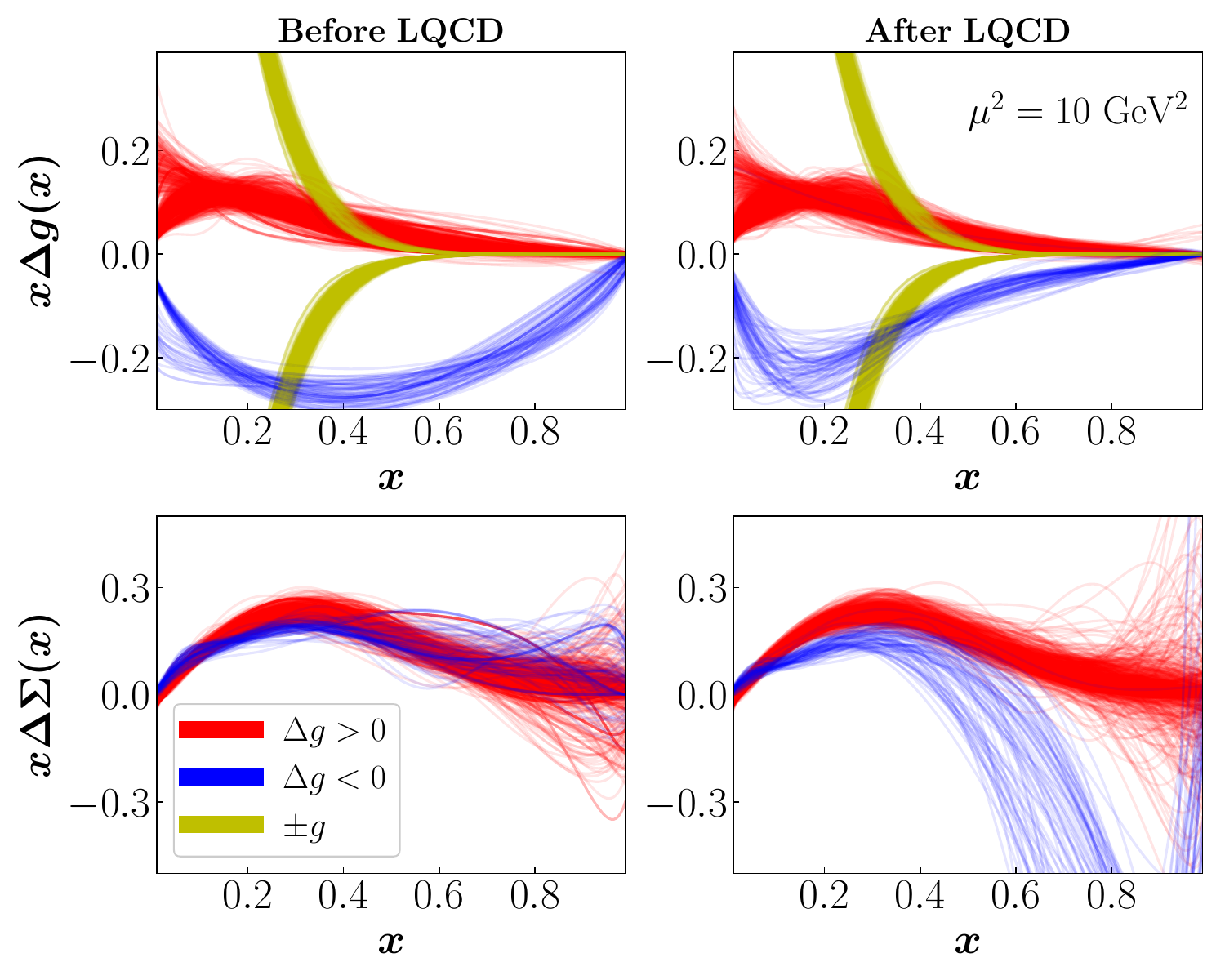}
    \caption{Impact of LQCD's Ioffe time pseudo-distributions on the gluon (upper) and quark singlet (lower) helicity distributions. The left column represents prior to lattice QCD data was included and the right represents after. }
    \label{f.hpdfs}
\end{figure*}

We now discuss the results at the hPDF level. In Fig.~\ref{f.hpdfs}, we present the replicas of $\Delta g$ and $\Delta \Sigma$ before and after the inclusion of LQCD data, categorizing the hPDFs by the sign of $\Delta g$. 
In the gluon sector, we observe significant changes for the negative solutions for $x>0.3$, where the behavior of the replicas tends to violate the positivity constraints less. Nevertheless, negativity in the gluon helicity is still visible in the region $x<0.2$, which cannot be ruled out by the positivity constraints or any of the present data from experiments or LQCD included in the present analysis. Interestingly, for the quark singlet sector, we find, in contrast to the no-LQCD case, differences in $\Delta \Sigma$ for $x>0.3$, where negative solutions appear which corresponds to negative $\Delta g$ solutions. As mentioned before, our DIS DSAs are in the region with $W^2>10~{\rm GeV^2}$ with the highest value of $x~\sim0.66$ hence insensitive to most of the negative $\Delta \Sigma$ above $x>0.7$ and in turn it prevents the DSAs from single jet productions to discriminate against the negative solutions of $\Delta g$.

\begin{figure*}[h!]
    \centering
    \includegraphics[width=\textwidth]{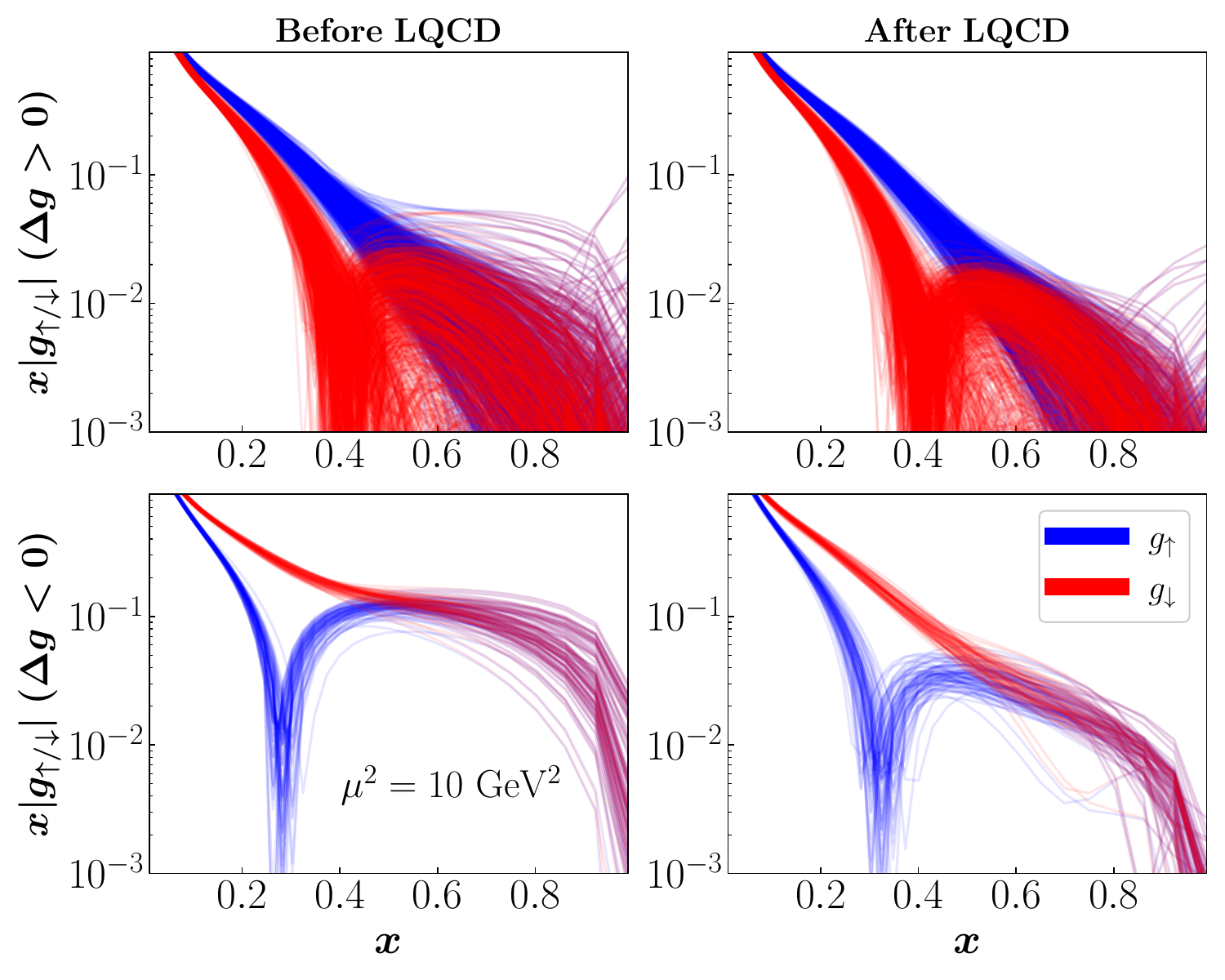}
    \caption{
    Impact of lattice QCD's Ioffe time distributions on the individual gluon helicity components in absolute value. The left and right columns shows the results prior and after to the inclusion of LQCD data respectively. }
    \label{f.abs_ga_vs_gna}
\end{figure*}

Finally, in Fig.~\ref{f.abs_ga_vs_gna}, we display the individual components of the gluon helicity PDF, namely $g_{\uparrow}$ and $g_{\downarrow}$. In the case of $\Delta g>0$, we observe violations of positivity, mostly for the spin anti-aligned PDF $g_{\downarrow}$, above $x\sim 0.4$. For the mirror version, $\Delta g<0$, this violation occurs earlier, around $x\sim 0.3$, for the spin-aligned PDF $g_{\uparrow}$. As mentioned before, positivity constraints are violated regardless of the sign of $\Delta g$.

\section{Conclusions}
\label{sec.conclusions}

We have performed a new global analysis of spin-dependent parton distribution functions, incorporating Ioffe time pseudo-distributions computed in lattice QCD, which directly probe the gluon helicity PDF. Our analysis critically examines the overall agreement between data and theory. We find that the inclusion of the LQCD data does not significantly alter the quality of the results. At present, LQCD data do not definitively rule out the negative $\Delta g$ solutions, which were recently found by the JAM collaboration at moderate values of $x$. Nevertheless, we observe changes in the shape and magnitude of the gluon helicity PDF and the quark sector. LQCD data reduces the magnitude of the negative $\Delta g$ solutions at high $x$, leading to a sign change in the corresponding quark singlet solutions at $x\sim 0.4$, necessary to describe the polarized jet data from RHIC.

The changes induced by LQCD data do not impact the description of inclusive DIS data extending up to $x\approx 0.66$. Future work should include the large-$x$ data from Jefferson Lab, which requires additional treatment of power corrections. However, these data are likely to exhibit tension with the negative $\Delta g$ and negative $\Delta \Sigma$ solutions at high $x$, providing an empirical test of the sign of $\Delta g$. Nevertheless, we emphasize the importance of including additional large-$x$ data that are less sensitive to power corrections in order to comprehensively assess the universality of the resulting hPDFs.

For future work, we look forward to incorporating dijet data from RHIC, which may help constrain the sign of $\Delta g$ at high $x$. The proposed JLab~24~GeV upgrade would also give greater discriminating power at larger $x$ values~\cite{Accardi:2023chb}. Furthermore, forthcoming LQCD calculations sensitive to the singlet distribution $\Delta \Sigma$ may provide new insights into the high-$x$ behavior of hPDFs. We should also note that this study is limited by the data currently available, and anticipate collecting additional crucial information from the future Electron-Ion Collider~\cite{Accardi:2012qut}, which is expected to provide constraints on hPDFs in the previously unexplored region of small $x$ and large $Q^2$, with observables that are sensitive linearly to $\Delta g$.

\acknowledgments
We would like to thank Werner Vogelsang for useful discussions.  This project was supported by the U.S. Department of Energy, Office of Science, Contract No.~DE-AC05-06OR23177, under which Jefferson Science Associates, LLC operates Jefferson Lab. N.S. was supported by the DOE, Office of Science, Office of Nuclear Physics in the Early Career Program. C.J.M.~is supported in part by the U.S.~DOE EC Award \mbox{\#DE-SC0023047}. S.Z.~acknowledges support by the French Centre national de la recherche scientifique (CNRS) under an Emergence@INP 2023 project. R.M.W. was supported by N.S.'s Early Career Award. KO was supported in part  by the U.S.~DOE Grant \mbox{\#DE-FG02-04ER41302}. 
This work has benefited from the collaboration enabled by the Quark-Gluon Tomography (QGT) Topical Collaboration, U.S.~DOE Award DE-SC0023646.
Computations for this work were carried out in part on facilities of the USQCD Collaboration, which are funded by the Office of Science of the U.S. Department of Energy. This work was performed in part using computing facilities at William and Mary which were provided by contributions from the National Science Foundation (MRI grant PHY-1626177), and the Commonwealth of Virginia Equipment Trust Fund. This work used the Extreme Science and Engineering Discovery Environment (XSEDE), which is supported by National Science Foundation grant number ACI-1548562. Specifically, it used the Bridges system, which is supported by NSF award number ACI-1445606, at the Pittsburgh Supercomputing Center (PSC) \cite{6866038, Nystrom:2015:BUF:2792745.2792775}. In addition, this work used resources at NERSC, a DOE Office of Science User Facility supported by the Office of Science of the U.S. Department of Energy under Contract \#DE-AC02-05CH11231, as well as resources of the Oak Ridge Leadership Computing Facility at the Oak Ridge National Laboratory, which is supported by the Office of Science of the U.S. Department of Energy under Contract No. \mbox{\#DE-AC05-00OR22725}. The software codes {\tt Chroma} \cite{Edwards:2004sx}, {\tt QUDA} \cite{Clark:2009wm, Babich:2010mu}, {\tt QPhiX} \cite{QPhiX2}, and {\tt Redstar} \cite{Chen:2023zyy} were used in our work. The authors acknowledge support from the U.S. Department of Energy, Office of Science, Office of Advanced Scientific Computing Research and Office of Nuclear Physics, Scientific Discovery through Advanced Computing (SciDAC) program, and of the U.S. Department of Energy Exascale Computing Project. The authors also acknowledge the Texas Advanced Computing Center (TACC) at The University of Texas at Austin for providing HPC resources, like Frontera computing system~\cite{10.1145/3311790.3396656} that has contributed to the research results reported within this paper.
We acknowledge PRACE (Partnership for Advanced Computing in Europe) for awarding us access to the high performance computing system Marconi100 at CINECA (Consorzio Interuniversitario per il Calcolo Automatico dell’Italia Nord-orientale) under the grants Pra21$-$5389 and Pra23$-$0076. This work also benefited from access to the Jean Zay supercomputer at the Institute for Development and Resources in Intensive Scientific Computing (IDRIS) in Orsay, France under project A0080511504.

\bibliography{bibfiles/expdata,bibfiles/jam,bibfiles/lqcd}

\end{document}